\documentclass[10pt, a4paper]{article}
\usepackage{lrec}
\usepackage{graphicx}
\usepackage{tabularx}
\usepackage{soul}
\usepackage{float}
\usepackage{epstopdf}
\usepackage[utf8]{inputenc}

\usepackage{dsfont}
\usepackage{amsmath}

\DeclareMathOperator{\desc}{\mathit{describe}}
\DeclareMathOperator{\id}{\mathit{identify}}

\usepackage{lipsum}

\usepackage{hyperref}
\usepackage{cleveref}
\usepackage{xstring}
\usepackage{listings}
\usepackage{xcolor}
\usepackage{markdown}
\usepackage[labelfont=bf]{caption}

\crefformat{footnote}{#2\footnotemark[#1]#3}

\colorlet{punct}{red!60!black}
\definecolor{background}{HTML}{EEEEEE}
\definecolor{delim}{RGB}{20,105,176}
\colorlet{numb}{magenta!60!black}

\lstdefinelanguage{json}{
    basicstyle=\normalfont\ttfamily,
    numbers=left,
    numberstyle=\scriptsize,
    stepnumber=1,
    numbersep=8pt,
    showstringspaces=false,
    breaklines=true,
    frame=lines,
    backgroundcolor=\color{background},
    literate=
     *{0}{{{\color{numb}0}}}{1}
      {1}{{{\color{numb}1}}}{1}
      {2}{{{\color{numb}2}}}{1}
      {3}{{{\color{numb}3}}}{1}
      {4}{{{\color{numb}4}}}{1}
      {5}{{{\color{numb}5}}}{1}
      {6}{{{\color{numb}6}}}{1}
      {7}{{{\color{numb}7}}}{1}
      {8}{{{\color{numb}8}}}{1}
      {9}{{{\color{numb}9}}}{1}
      {:}{{{\color{punct}{:}}}}{1}
      {,}{{{\color{punct}{,}}}}{1}
      {\{}{{{\color{delim}{\{}}}}{1}
      {\}}{{{\color{delim}{\}}}}}{1}
      {[}{{{\color{delim}{[}}}}{1}
      {]}{{{\color{delim}{]}}}}{1},
}

\title{WIKIR: A Python toolkit for building a large-scale Wikipedia-based\\ English Information Retrieval Dataset}

\name{Jibril Frej, Didier Schwab, Jean-Pierre Chevallet}

\address{Univ. Grenoble Alpes, CNRS, Grenoble INP*, LIG, 38000 Grenoble, France\\
* Institute of Engineering Univ. Grenoble Alpes\\
        \{jibril.frej, didier.schwab, jean-pierre.chevallet\}@univ-grenoble-alpes.fr\\}

\abstract{
Over the past years, deep learning methods allowed for new state-of-the-art results in \textit{ad-hoc} information retrieval. However such methods usually require large amounts of annotated data to be effective. Since most standard \textit{ad-hoc} information retrieval datasets publicly available for academic research (e.g. \textit{Robust04}, \textit{ClueWeb09}) have at most 250 annotated queries, the recent deep learning models for information retrieval perform poorly on these datasets. These models (e.g. DUET, Conv-KNRM) are trained and evaluated on data collected from commercial search engines not publicly available for academic research which is a problem for reproducibility and the advancement of research. In this paper, we propose \textit{WIKIR}: an open-source toolkit to automatically build large-scale English information retrieval datasets based on \textit{Wikipedia}. \textit{WIKIR} is publicly available on \textit{GitHub}. We also provide wikIR78k and wikIRS78k: two large-scale publicly available datasets that both contain 78,628 queries and 3,060,191 (query, relevant documents) pairs.\newline  }

\begin{document}

\maketitleabstract

\section{Introduction}

Deep learning has been shown to be effective in various natural language processing (NLP) tasks such as language modeling, reading comprehension, question answering and natural language understanding~\cite{devlin-etal-2019-bert,DBLP:journals/corr/abs-1906-08237}. However, both large and public datasets are key factors for developing effective and reproducible deep learning models.

\textit{Ad-hoc} information retrieval (IR) consists in ranking a set of unstructured documents with respect to a query. Despite the progress in NLP using deep neural networks (DNNs), \textit{ad-hoc} IR on text documents has not benefited as much as other fields of NLP from DNNs yet~\cite{46572}. The absence of significant success in \textit{ad-hoc} IR using deep learning approaches is mainly due to the complexity of solving the ranking task using only unlabelled data~\cite{46572}. Consequently, the availability of large amount of labelled data is crucial to develop effective DNNs for \textit{ad-hoc} IR. However, as described in Table~\ref{tab:stats}, most of the publicly available English IR datasets only have few labelled data with at most 1,692 labelled queries. 

Other datasets than the ones presented in Table~\ref{tab:stats}, such as \textit{Yahoo! LETOR}~\cite{chapelle2011yahoo}, with more labelled data ($\approx$30k labelled queries) are publicly available. However, only the feature vectors describing query-document pairs are provided. Such datasets are suitable for feature-based learning-to-rank models but not for DNNs that require the original content of queries and documents.

Thus, most of the deep learning model for \textit{ad-hoc} IR that have been proposed recently are developed using one of the following approaches:

\textbf{(1)} Using large amounts of data collected from commercial search engines that are not publicly available~\cite{DBLP:journals/corr/abs-1904-09171,Mitra:2017:LMU:3038912.3052579}. This process is expensive, time consuming and not reproducible.

\textbf{(2)} Using publicly available datasets that have few annotated data such as \textit{MQ2007} and \textit{MQ2008}~\cite{Pang:2017:DND:3132847.3132914,Fan:2018:MDR:3209978.3209980}. This approach can restrain the model design due to the lack of data.

\textbf{(3)} Using weak supervision that consists in pre-training a supervised model on data labelled with an unsupervised approach~\cite{46572}. However, this method can bias large models to rank similarly as the unsupervised ranker.

Recently, \newcite{Zheng:2018:SND:3209978.3210092} proposed \textit{Sogou-QCL}, a publicly available dataset in Chinese with click relevance label. To the best of our knowledge, \textit{Sogou-QCL} is the only public large-scale ($\approx$500k queries) dataset for \textit{ad-hoc} IR. The release of this dataset was the first step in reproducible research on neural ranking model applied to \textit{ad-hoc} IR.

\textit{Wikipedia} has recently been used to build large-scale cross-lingual information retrieval (CLIR) datasets to train effective neural learning-to-rank models~\cite{schamoni2014learning}.

Leveraging this idea, we propose \textit{WIKIR}: a toolkit to build a \textit{Wikipedia}-based large-scale English IR dataset.
\textit{WIKIR} can also be used to train and evaluate several deep text matching models on the datasets it created.

Moreover, we propose a general framework to build IR datasets automatically from any set of documents constrained by three topical properties that will be introduced further (see Section~\ref{subsec:prop}).


Our contributions are fourfold:

\begin{itemize}
    \item We provide \textit{WIKIR}: a toolkit\footnote{\url{https://github.com/getalp/wikIR}} to build a \textit{Wikipedia}-based English Information Retrieval dataset;
    \item We present a framework for creating IR datasets from a set of documents that satisfies three topical properties: \textit{Existence}, \textit{Identifiability} and \textit{Describability};
    \item We propose wikIR78k and wikIRS78k: two large-scale datasets generated with \textit{WIKIR}, publicly available for download\footnote{\url{https://www.zenodo.org/record/3707606}}\footnote{\url{https://www.zenodo.org/record/3707238}};
    \item We provide \textit{Python} scripts to train and evaluate deep learning models for \textit{ad-hoc} IR on our datasets.
\end{itemize}

\begin{table}
\centering
\begin{tabular}{c|c|c|c}
    \hline
    Dataset & \#Query & \#Doc & Avg \#$d^+/q$\\
    \hline
    CLEF 2014 & 50 & 1M & 64.56\\
    ClueWeb09 & 200 & 1B & 74.62 \\
    ClueWeb12 & 100 & 733M & 189.63 \\
    GOV2 & 150 & 25M & 181.51 \\
    MQ2007 & 1,692 & 65k & 10.63 \\
    MQ2008 & 784 & 14k & 3.82 \\
    Robust04 & 250 & 0.5M & 63.28 \\
\hline
\end{tabular}
\caption{Statistics of several publicly available English IR Dataset where the original query and document contents are available. Avg \#$d^+/q$ denotes the average number of relevant document per query.}
\label{tab:stats}
\end{table}

\section{A general framework for automatic IR dataset creation}

In this section, we propose a general framework to create automatically an IR dataset from a resource $\mathcal{R}$ composed of a set of documents. An IR dataset is composed of:
\begin{itemize}
    \item[-] $\mathcal{D}$, a set of documents;
    \item[-] $\mathcal{Q}$, a set of queries;
    \item[-] $\mathcal{R}el$, a set of relevance labels for each query-document pairs~\cite{schutze2008introduction}.
\end{itemize}

\subsection{Properties}\label{subsec:prop}

We define 3 properties that $\mathcal{R}$ must satisfy to be used to build an IR dataset.
\vspace{0.5em}

\textbf{Topical Existence.} There exists at least one topic related to each document in $\mathcal{R}$.\\
\textit{Topical Existence} guarantees the topical relevance~\cite{mizzaro1997relevance} of documents with respect to a subject.
\vspace{0.5em}

\textbf{Topical Identifiability.} There exists a function $\id()$ that identifies all the topics related to any document of $\mathcal{R}$.\\
Using \textit{Topical Identifiability}, we can assess the relevance of documents with respect to the topics in $\mathcal{R}$.
\vspace{0.5em}

\textbf{Topical Describability.} There exists a function $\desc()$ that associates every topic with a short and accurate description.\\
\textit{Topical Describability} is desirable to be able to build queries from the topics in the resource $\mathcal{R}$.

\subsection{Dataset construction}\label{coll_constr}

In the following, we describe how to use a resource $\mathcal{R}$ that satisfies the three properties listed above to automatically construct an IR dataset.

\textbf{Document construction.} We choose a subset of the resource $\mathcal{R}$ to construct the set of documents: $\mathcal{D} \subseteq \mathcal{R}$. For example, if $\mathcal{R}$ is the set of \textit{Wikipedia} articles, we can choose $\mathcal{D}$ to be the set of \textit{Wikipedia} articles that contain more than 1000 words.

\textbf{Query construction.} We start by identifying all topics in the set of documents $\mathcal{D}$ using the $\id()$ function:
\begin{equation*}
    \mathcal{T}_{\mathcal{D}} = \bigcup\limits_{d \in \mathcal{D}} \id(d),
\end{equation*}
where $\mathcal{T}_{\mathcal{D}}$ is the set of all topics in $\mathcal{D}$. 
Then, we use the $describe()$ function on all of the topic to construct the query set $\mathcal{Q}$: 

\begin{equation*}
    \mathcal{Q} = \left\{ describe(t) \big| t \in \mathcal{T}_{\mathcal{D}} \right\}.
\end{equation*}

\textbf{Relevance label construction.} $\mathcal{R}el$ is the set of all (query-document-relevance) triplets:

\begin{equation*}
    \mathcal{R}el = \left\{ \big(q,d,rel(q,d) \big) \big| \big(q,d \big) \in \mathcal{Q} \times \mathcal{D} \right\},
\end{equation*}

with $rel()$ a function that associates every query-document pairs with a relevance label. We propose to assign a positive relevance label (denoted $val^+$) to document $d$ with respect to query $q$ if $d$ contains the topic that was used to build~$q$. Otherwise a negative or null relevance label (denoted $val^-$) is assigned:

\begin{equation*}
    rel(q,d) = 
    \begin{cases} 
        val^+ \in \mathds{R}^+_*,&\mbox{if } t_q \in \id(d),\\
        val^- \in \mathds{R}^-,&\mbox{else},
    \end{cases}
\end{equation*}

where $t_q$ stands for the topic that was used to build query $q$: $\desc(t_q) = q$

\subsection{The case of \textit{Wikipedia}}\label{subsec:wikicase}
\begin{figure*}
    \includegraphics[width=\linewidth]{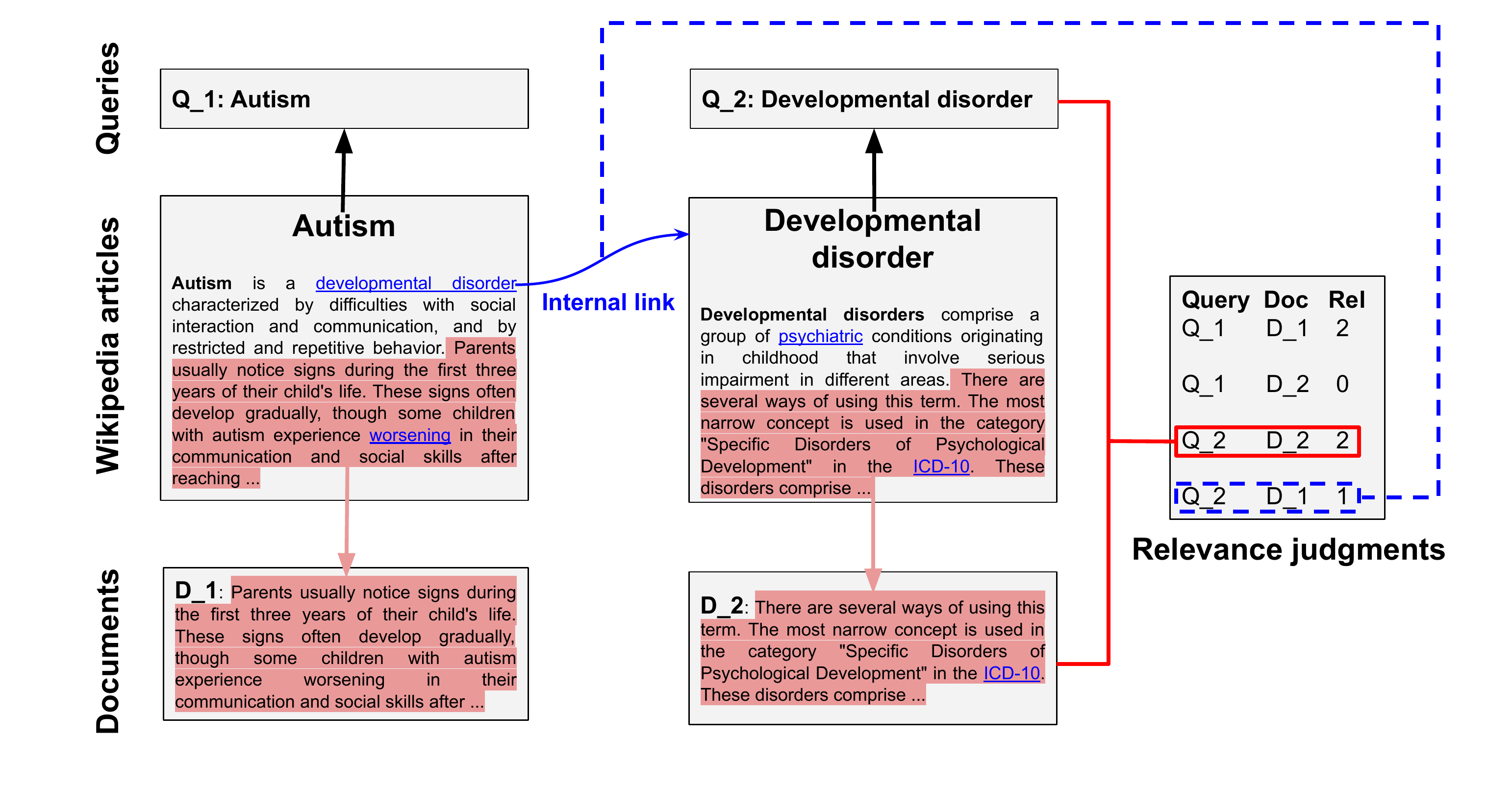}
    \caption{Description of the construction process of an IR dataset by \textit{WIKIR} using only two articles. Queries are built from the title of articles. Documents are constructed using the full text of articles without the title and without the first sentence. A relevance label equal to 2 is assigned to query and documents that are built from the same article. A relevance label equal to 1 is assigned using internal links in the first sentence of articles.}
    \label{fig:wikir_descr}
\end{figure*}

In this subsection we show that the set of English \textit{Wikipedia} articles $W$ does satisfy \textit{Topical Existence}, \textit{Describability} and \textit{Identifiability}. A simplified description of the construction process of an IR dataset using 2 articles from \textit{Wikipedia} is displayed in Figure~\ref{fig:wikir_descr}.


\textbf{Topical Existence.} Every \textit{Wikipedia} article is related to at least one topic: its main subject.\\
\textbf{Topical Identifiability.} We assume that if an article $a$ contains an internal link to another article $a_t$ in its first sentence (denoted $f_a$), then the main subject of $a_t$ is a topic of $a$. The intuition behind this assumption is that the first sentence of most \textit{Wikipedia} articles is a good description of the article's content~\cite{sasaki-etal-2018-cross} and if a link is present, it points to an important topic of the considered article. Therefore, we propose to define $\id()$ as follows:
\begin{equation}
    \id(a) = \left\{s_a\right\} \bigcup \left\{ s_{a_t} \in \mathcal{W} \big| \exists \; f_{a_t} \xrightarrow[\text{link}]{} a \right\},
    \label{eq:id}
\end{equation}
where $s_a$ denotes the main subject of article $a$ and $f_{a_t} \xrightarrow[\text{link}]{} a$ designs an internal link in the first sentence of article $a_t$ that points to article $a$. Thus, $\id()$ considers the set of topics related to article $a$ as the main subject of $a$: $s_a$  and the main subject of all articles that points to $a$ in their first sentence. For example, the set of topics related to the article \textit{Developmental disorder} is its main subject and the main subject of the article \textit{Autism} because there is a link in the first sentence of article \textit{Autism} that points to article \textit{Developmental disorder} (see Figure~\ref{fig:wikir_descr}).

\textbf{Topical Describability.} Because topics are main subjects of \textit{Wikipedia} articles, one way to get a short and accurate description is to use the article title: 

\begin{equation}
    \desc(s_a) = title_a,
    \label{eq:desc}
\end{equation}

where $title_a$ is the title of article $a$. To get a long and noisy topic description, we can also use the article first sentence: 

\begin{equation}
    \desc(s_a) = f_{a_t}.
    \label{eq:desc_sent}
\end{equation}

\section{WIKIR toolkit description}\label{WIKIR}
In this section, we describe \textit{WIKIR} toolkit and make explicit the motivations behind some design decisions. For an exhaustive list of the options available and to have examples on how to use \textit{WIKIR}, please check our \textit{github} repository: \url{https://github.com/getalp/wikIR} \\
\subsection{WIKIR for dataset creation}
To create a dataset using an XML \textit{Wikipedia} dump file from \textit{Wikimedia} database backup dumps.\footnote{\url{https://dumps.wikimedia.org/backup-index.html}} \textit{WIKIR} follows 3 main steps: construction, processing and storing.

\subsubsection{Dataset construction}\label{WIKIR_creat}

\textbf{Wikipedia dump extraction.} We use \textit{WikiExtractor}\footnote{\url{https://github.com/attardi/wikiextractor}} to extract plain text from an English \textit{Wikipedia} dump. We end up with a \textit{json} file (described in Figure~\ref{fig:my_label}) that contains the URL, title and text of all \textit{Wikipedia} articles. When using \textit{wikiextractor}, we use the option to preserve links in the text in order to build relevance labels.\\
\textbf{Document extraction.} The set of documents $\mathcal{D}$ is extracted using the ``text" field associated to each article in the \textit{json} file produced by the previous step. The first line of the ``text" field (that corresponds to the article title) is deleted. We also remove article title from documents in order to avoid the following situation: given a query, the most relevant document will always starts with the query itself which makes the ranking task significantly easier.\\
\textbf{Query construction.} As described in Section~\ref{coll_constr} to build queries we need an $\id()$ function and a $\desc()$ function. \textit{WIKIR} uses the $\id()$ function defined in equation~(\ref{eq:id}). The $\desc()$ function is defined using equation~(\ref{eq:desc}) or equation~(\ref{eq:desc_sent}). To sum up, topics are identified using internal links and are described using either article titles or article first sentences. The construction process of queries is the same as in Section~\ref{coll_constr}

\textbf{Relevance label construction.} As explained in Section~\ref{coll_constr} in order to build $\mathcal{R}el$ we need to define $rel()$. To do so, we assume that the most relevant document for a query is the document built from the same article as the query. Consequently, we define $rel()$ as:

\begin{equation*}
    rel(q,d) = 
    \begin{cases} 
        2, &\mbox{if } a_q = a_d,\\ 
        1, &\mbox{if } a_d \in \id(d) \setminus {a_d},\\
        0 ,&\mbox{otherwise}, 
    \end{cases}
\end{equation*}

where $a_q$  (resp. $a_d$) denotes the \textit{Wikipedia} article used to build query $q$ (resp. document $d$). Thus we assign a relevance label equal to two for query-document pairs that come from the same article. We assign a relevance label equal to one to a query-document pair if there is a link from the first sentence of the article of the document that points to the article of the query. For example, if we consider the query \textit{``Developmental disorder"}, the most relevant (relevance~=~2) document is \textit{``Developmental disorders comprise a group of \ldots"} because they are built from the same the article. The document \textit{``Autism is a developmental disorder characterized by \ldots"} is relevant (relevance~=~1) because the article \textit{Autism} contains a link to the \textit{Developmental disorder} article (see Figure~\ref{fig:wikir_descr}).

\subsubsection{Dataset processing}

\textbf{Query selection.} In order to have a balanced dataset, we select only queries that have a minimum number of relevant documents (5 by default). We also limited queries length to a maximum of 10 words.\\
\textbf{Preprocessing.} \textit{WIKIR} starts by deleting the target in hypertext references (\textit{href}) but keeps the text. For example,\\ \textit{``\textless
a href=\textbackslash ``Regressive\%20autism\textbackslash"\textgreater worsening\textless/a\textgreater"} becomes \textit{``worsening"}. Then, every non alphanumerical character is deleted. By default \textit{WIKIR} also lowercases all the characters in the dataset.\\
\textbf{Separation into training, validation and test sets.} Queries and their corresponding relevance label (\textit{qrels}) are randomly separated into training, validation and test sets. Documents are not separated as well because in \textit{ad-hoc} IR, we assume to have a fixed set of documents to retrieve from~\cite{ribeiro1999modern}.



\begin{figure*}
    \begin{lstlisting}[language=json,firstnumber=1]
    {"id": "12",
     "url": "https://en.wikipedia.org/wiki?curid=12",
     "title": "Anarchism",
     "text": "Anarchism\n\nAnarchism is an <a href=\"anti-authoritarian\">anti-authoritarian</a> <a href=\"political%20philosophy\">political philosophy</a> that advocates ... "
    }
    {"id": "25",
     "url": "https://en.wikipedia.org/wiki?curid=25",
     "title": "Autism",
     "text": ""Autism\n\nAutism is a <a href=\"developmental%20disorder\">developmental disorder</a> characterized by difficulties with ... "
    }
    ...
\end{lstlisting}
    \caption{\textit{json} file extracted from English \textit{Wikipedia} dump using \textit{WikiExtractor}}
    \label{fig:my_label}
\end{figure*}

\subsection{WIKIR for BM25: a first stage ranker}\label{sebsec:bm25}
\subsubsection{Motivation}
After the dataset is created, \textit{WIKIR} can be used to run Okapi BM25~\cite{Robertson:1994:SEA:188490.188561}: a state-of-the art IR model compatible with an inverted index. An inverted index is a structure to store the documents of an IR dataset that makes the retrieval of documents extremely efficient~\cite{DBLP:journals/nle/Sanderson10}. We propose this option because the vast majority of DNNs developed for \textit{ad-hoc} IR are not compatible with an inverted index~\cite{zamani2018neural}. They rely on a first ranking stage made by an efficient model such as BM25 and only re-rank the top-$k$ documents for a given query in order to have an efficient search. Thus \textit{WIKIR} can be used to run BM25 and save the top-$k$ documents for each query.

\subsubsection{Implementation}
Instead of using a common information retrieval system (IRS) such as \textit{Terrier},\footnote{\label{terrier}\url{http://terrier.org/}} \textit{Lucene}\footnote{\url{http://lucene.apache.org/}} or \textit{Lemur}\footnote{\url{http://www.lemurproject.org}} to run and evaluate BM25 on our dataset, we used the \textit{Python} library \textit{Rank-BM25}.\footnote{\url{https://github.com/dorianbrown/rank_bm25}} We made this decision to facilitate the use of \textit{WIKIR} and to aid the reproducibility of our experiments that do not require the installation of any software that is not in our \textit{GitHub} repository. Because \textit{Rank-BM25} does not preprocess text, we used \textit{nltk} \textit{Python} library~\cite{DBLP:journals/corr/cs-CL-0205028} to apply Porter stemmer~\cite{porter2001snowball} and stopword removal as commonly done in IR. It should be noted that we applied stemming and stopword removal only for BM25: the queries and documents in the dataset created by \textit{WIKIR} are not stemmed and do contain stopwords.


\subsection{WIKIR for neural re-ranking}

\textit{WIKIR} can be used to train and evaluate DNNs on the dataset it created. As explained in Section~\ref{sebsec:bm25}, we perform neural re-ranking using BM25 as a first stage ranker. We used \textit{MatchZoo} deep text matching library for training and evaluation of the models. We used \textit{MatchZoo} because it has been accepted as a reliable toolkit for deep text matching research~\cite{Guo:2019:MLP:3331184.3331403}. Any model available in \textit{MatchZoo} can be trained and evaluated with \textit{WIKIR}. Once the training is done and the rankings of documents are saved, our toolkit can be used to compute evaluation measures, statistical significance and display the performance of each model in a format compatible with a \LaTeX \ table.


\begin{table}
\centering
\begin{tabular}{lcc}
     & wikIRS78k & wikIR78k \\
    \hline
    Document count   & 2.4M  & 2.4M \\
    Average document length & 744.58 & 744.58 \\
    Query count & 78k & 78k\\
    Average query length & 2.45 & 9.80\\
    Avg \#$d^+/q$ & 39.02 & 39.02\\
\hline
\end{tabular}
\caption{Statistics of wikIR78k and wikIRS78k. Avg \#$d^+/q$ denotes the average number of relevant document per query.}
\label{tab:mystats}
\end{table}

\section{Datasets}
In this section, we describe wikIR78k and wikIRS78k: the two datasets created by \textit{WIKIR} that we used in our experiments.\\
\textbf{wikIR78k.} wikIR78k is a large-scale dataset that contains 78,631 annotated queries. To build wikIR78k, we used the full set of \textit{Wikipedia} articles. To build queries, we used article titles. Moreover, we deleted the first sentence of each article when constructing the documents. We made this choice since all the information we use to assess relevance is contained in the first sentence of articles (see Section~\ref{coll_constr}) and we do not want DNNs that take into account word order to use this bias to their advantage. \\
\textbf{wikIRS78k.} The construction process of wikIRS78k is the same as wikIR78k, with the exception of  queries construction: we used articles first sentences instead of article titles. We propose a dataset with short and well defined queries and a dataset with long and noisy queries to study the robustness of IR models against noisy queries. Statistics of the datasets are displayed on Table~\ref{tab:mystats}. Queries are randomly split into training, validation and tests sets of size 80\% ,10\% ,10\% respectively.\\

\begin{table*}
\centering
\begin{tabular}{lllllllll}
\multicolumn{9}{c}{\large\textbf{wikIR78k}}\\
\hline
Model & P@5 & P@10 & P@20 & nDCG@5 & nDCG@10 & nDCG@20 & nDCG & MAP\\
\hline
\hline
BM25 & 0.2622 & 0.2039 & 0.1498 & 0.3269 & 0.3045 & 0.3098 & 0.3555 & 0.1498 \\
\hline
ArcI & 0.1412\textsuperscript{\textbf{-}} & 0.1316\textsuperscript{\textbf{-}} & 0.1171\textsuperscript{\textbf{-}} & 0.1393\textsuperscript{\textbf{-}} & 0.1510\textsuperscript{\textbf{-}} & 0.1749\textsuperscript{\textbf{-}} & 0.2537\textsuperscript{\textbf{-}} & 0.0841\textsuperscript{\textbf{-}} \\
ArcII & 0.1492\textsuperscript{\textbf{-}} & 0.1401\textsuperscript{\textbf{-}} & 0.1224\textsuperscript{\textbf{-}} & 0.1428\textsuperscript{\textbf{-}} & 0.1559\textsuperscript{\textbf{-}} & 0.1799\textsuperscript{\textbf{-}} & 0.2560\textsuperscript{\textbf{-}} & 0.0885\textsuperscript{\textbf{-}} \\
MatchPyramid & 0.2302\textsuperscript{\textbf{-}} & 0.1886\textsuperscript{\textbf{-}} & 0.1485 & 0.2568\textsuperscript{\textbf{-}} & 0.2495\textsuperscript{\textbf{-}} & 0.2644\textsuperscript{\textbf{-}} & 0.3160\textsuperscript{\textbf{-}} & 0.1253\textsuperscript{\textbf{-}} \\
\hline
KNRM & 0.1288\textsuperscript{\textbf{-}} & 0.1199\textsuperscript{\textbf{-}} & 0.1078\textsuperscript{\textbf{-}} & 0.1186\textsuperscript{\textbf{-}} & 0.1296\textsuperscript{\textbf{-}} & 0.1531\textsuperscript{\textbf{-}} & 0.2402\textsuperscript{\textbf{-}} & 0.0761\textsuperscript{\textbf{-}} \\
DUET & 0.2645 & 0.2038 & 0.1533\textsuperscript{\textbf{+}} & 0.3323 & 0.3044 & 0.3082 & 0.3533 & 0.1447\textsuperscript{\textbf{-}} \\
DRMM & \textbf{0.2760}\textsuperscript{\textbf{+}} & \textbf{0.2122}\textsuperscript{\textbf{+}} & 0.1548\textsuperscript{\textbf{+}} & \textbf{0.3462}\textsuperscript{\textbf{+}} & \textbf{0.3189}\textsuperscript{\textbf{+}} & \textbf{0.3227}\textsuperscript{\textbf{+}} & \textbf{0.3653}\textsuperscript{\textbf{+}} & \textbf{0.1566}\textsuperscript{\textbf{+}} \\
Conv-KNRM & 0.2602 & 0.2057 & \textbf{0.1566}\textsuperscript{\textbf{+}} & 0.3080\textsuperscript{\textbf{-}} & 0.2906\textsuperscript{\textbf{-}} & 0.2992\textsuperscript{\textbf{-}} & 0.3422\textsuperscript{\textbf{-}} & 0.1419\textsuperscript{\textbf{-}} \\
\hline

&&&&&&&&\\
\multicolumn{9}{c}{\large\textbf{wikIRS78k}}\\
\hline
Model & P@5 & P@10 & P@20 & nDCG@5 & nDCG@10 & nDCG@20 & nDCG & MAP\\
\hline
\hline

BM25 & 0.2177 & 0.1634 & 0.1186 & 0.2944 & 0.2673 & 0.2695 & 0.3085 & 0.1163 \\
\hline
ArcI & 0.1156\textsuperscript{\textbf{-}} & 0.1076\textsuperscript{\textbf{-}} & 0.0953\textsuperscript{\textbf{-}} & 0.1096\textsuperscript{\textbf{-}} & 0.1201\textsuperscript{\textbf{-}} & 0.1418\textsuperscript{\textbf{-}} & 0.2104\textsuperscript{\textbf{-}} & 0.0650\textsuperscript{\textbf{-}} \\
ArcII & 0.1360\textsuperscript{\textbf{-}} & 0.1236\textsuperscript{\textbf{-}} & 0.1055\textsuperscript{\textbf{-}} & 0.1299\textsuperscript{\textbf{-}} & 0.1397\textsuperscript{\textbf{-}} & 0.1602\textsuperscript{\textbf{-}} & 0.2210\textsuperscript{\textbf{-}} & 0.0726\textsuperscript{\textbf{-}} \\
MatchPyramid & 0.2053\textsuperscript{\textbf{-}} & 0.1665 & 0.1271\textsuperscript{\textbf{+}} & 0.2296\textsuperscript{\textbf{-}} & 0.2232\textsuperscript{\textbf{-}} & 0.2336\textsuperscript{\textbf{-}} & 0.2722\textsuperscript{\textbf{-}} & 0.1025\textsuperscript{\textbf{-}} \\
\hline
KNRM & 0.1443\textsuperscript{\textbf{-}} & 0.1239\textsuperscript{\textbf{-}} & 0.1010\textsuperscript{\textbf{-}} & 0.1501\textsuperscript{\textbf{-}} & 0.1541\textsuperscript{\textbf{-}} & 0.1705\textsuperscript{\textbf{-}} & 0.2315\textsuperscript{\textbf{-}} & 0.0758\textsuperscript{\textbf{-}} \\
DUET & 0.2534\textsuperscript{\textbf{+}} & 0.1926\textsuperscript{\textbf{+}} & 0.1387\textsuperscript{\textbf{+}} & 0.3252\textsuperscript{\textbf{+}} & 0.2964\textsuperscript{\textbf{+}} & 0.2951\textsuperscript{\textbf{+}} & 0.3207\textsuperscript{\textbf{+}} & 0.1294\textsuperscript{\textbf{+}} \\
DRMM & 0.2368\textsuperscript{\textbf{+}} & 0.1769\textsuperscript{\textbf{+}} & 0.1275\textsuperscript{\textbf{+}} & 0.3188\textsuperscript{\textbf{+}} & 0.2872\textsuperscript{\textbf{+}} & 0.2868\textsuperscript{\textbf{+}} & 0.3197\textsuperscript{\textbf{+}} & 0.1248\textsuperscript{\textbf{+}} \\
Conv-KNRM & \textbf{0.2661}\textsuperscript{\textbf{+}} & \textbf{0.2026}\textsuperscript{\textbf{+}} & \textbf{0.1458}\textsuperscript{\textbf{+}} & \textbf{0.3253}\textsuperscript{\textbf{+}} & \textbf{0.3004}\textsuperscript{\textbf{+}} & \textbf{0.3010}\textsuperscript{\textbf{+}} & \textbf{0.3223}\textsuperscript{\textbf{+}} & \textbf{0.1351}\textsuperscript{\textbf{+}} \\
\hline
\end{tabular}
\caption{Performance comparison of different models on wikIR78k and wikIRS78k. Significant improvement/degradation with respect to BM25 is denoted as (+/-) with p-value $<$ 0.01.}
\label{table:res}
\end{table*}

\section{Experimental settings}
This section describes the experiments we conducted on our datasets.

\subsection{Models description}
We evaluated 3 types of models: bag-of-words, DNNs for text matching and DNNs for \textit{ad-hoc} IR.
\subsubsection{Exact matching model}
We use Okapi BM25: a state-of-the-art ranking function that uses exact matches between query and document terms~\cite{Robertson:1994:SEA:188490.188561}:

\begin{equation}
    \text{BM25}(q,d) = \sum\limits_{t \in q} \text{idf}_t \frac{\text{tf}_{td}(k_1 + 1)}{\text{tf}_{td} + k_1\left(1-b + b\frac{|d|}{avgdl}\right)}, 
\end{equation}

where q is a query, d is a document, $\text{tf}_{td}$ is the term frequency (number of occurrences) of term t in document d, $k_1$ and $b$ are hyperparameters of BM25 and $avgdl$ denotes the average length of documents in $C$. The inverse document frequency of term t denoted as $\text{idf}_t$ reflects the discriminative power of term t to assess relevance~\cite{schutze2008introduction}:

\begin{equation}
\text{idf}_t = \log\frac{|C|+1}{\text{df}_t},
\end{equation}

where $C$ is the considered collection of documents and $\text{df}_t$ is the document frequency of term t: the number of documents that contain term t.

\subsubsection{Deep neural networks for text matching}
Text matching is a general task that consists in computing a matching score between two texts. Models developed for text matching do not take into account IR specificities such as query term importance or exact matching signals consideration~\cite{Guo:2016:DRM:2983323.2983769}.\\
\textbf{ArcI.} A representation model that uses 1D-convolutions and pooling layers to get a fixed size representation of sentences. The similarity score is obtained with a multilayer perceptron (MLP) on the representations of the two inputs~\cite{NIPS2014_5550}.\\
\textbf{ArcII.} An interaction model that uses 1D-convolutions to build an interaction matrix of the two input sentences. The final score is obtained using 2D-convolutions, max-pooling and MLP on the interaction matrix~\cite{NIPS2014_5550}.\\
\textbf{MatchPyramid.} An interaction model that build an interaction matrix between the two input sentences using the dot product between their word embeddings. The matrix obtained is processed using a convolutional neural network (CNN) and the matching score is computed using a MLP on the output of the CNN~\cite{pang2016text}.

\subsubsection{Deep neural networks for \textit{ad-hoc} IR}

\textbf{DRMM.} Uses a matching histogram between query term and all of the document terms, followed by a MLP to get a query term score. The final matching score is the sum of all query terms scores~\cite{Guo:2016:DRM:2983323.2983769}.\\
\textbf{KNRM.} A neural ranking model that uses word interactions and kernel pooling to produce learning-to-rank features. The final score is computed with a linear layer and a non-linear activation function applied on the ranking features~\cite{Xiong:2017:ENA:3077136.3080809}.\\
\textbf{DUET.} Model that uses both local (exact matching of n-grams of characters) and distributed (word embeddings) representations to compute a relevance score~\cite{Mitra:2017:LMU:3038912.3052579}.\\
\textbf{Conv-KNRM.} As KNRM, Conv-KNRM~\cite{Dai:2018:CNN:3159652.3159659} is based on kernel pooling to produce learning-to-rank features but it uses convolutions to match n-grams of words and has multiple interaction matrices.

\subsection{Implementation details}

\textbf{Training.}   
Each training sample consists of a query $q$, a document $d^+$ relevant to $q$ and a set of 5 irrelevant documents $D^-$ with respect to $q$. We use the cross entropy loss function for ranking provided by \textit{MatchZoo} defined as:

\begin{equation*}
    \mathcal{L}(q,d^+,D^-) = rel(q,d^+) \log  \frac{\exp\left(s(q,d^+)\right)}{\sum\limits_{d^- \in D^-} \exp\left(s(q,d^-)\right)}
\end{equation*}
where $s(q,d)$ denoted the score of $d$ with respect to $q$. We used the cross entropy loss function for ranking instead of the widely used Hinge loss function for pairwise training of \textit{ad-hoc} IR models~\cite{GUO2019102067} as preliminary experiments showed that the cross entropy loss function is more efficient in terms of training time and produces more effective models. We use the Adam optimizer~\cite{DBLP:journals/corr/KingmaB14} with a learning rate equals to 0.001. Each model is trained 5 times (with different initialization) for 50 epochs. We select the model that has the highest normalized discounted cumulative gain~\cite{Jarvelin:2002:CGE:582415.582418} on the validation set and report its results on the test set.
\textbf{Embeddings.} We used Glove~\cite{pennington2014glove} word embeddings of dimension 300 provided by \textit{MatchZoo}.\\
\textbf{Hyperparameters.} BM25 hyperparameters are set to their default values in \textit{Rank-BM25}: $k_1=1.5$ and $b=0.75$. Hyperparameters associated with DNNs (e.g., number of layers, kernel size, similarity function) were set to their default value implemented in \textit{MatchZoo}, except for the dropout rate that we set to 0.5 for models with a dropout parameter.\\
\textbf{Evaluation metrics.} We use 3 standard evaluation metrics: MAP, Precision and normalized discounted cumulative gain (nDCG). We use a two-tailed paired t-test with Bonferroni correction to measure statistically significant differences between the evaluation metrics~\cite{DBLP:conf/sigir/UrbanoMM13a,Fuhr:2018:CMI:3190580.3190586}.

\section{Results and discussion}

\subsection{Short and well defined queries}
As we can see on Table~\ref{table:res}, when queries are short and well defined (wikIR78k) BM25 is a strong baseline. Indeed, only the DRMM model manages to outperform BM25 on all metrics with statistical significance. Moreover, even though the DUET and Conv-KNRM models were designed for \textit{ad-hoc} IR, they do not manage to outperform BM25.

Models that were not designed for \textit{ad-hoc} IR but for text matching perform statistically significantly worst than BM25. This suggests that datasets created with \textit{WIKIR} are suited for designing and training DNNs specifically for \textit{ad-hoc} IR.\\

\subsection{Long and noisy queries}
Interestingly, models react differently to noisy queries (wikIRS78k). BM25 and DRMM are strongly affected by noise ($-9.94\%$ and $-7.91\%$, respectively on the nDCG@5 compared to wikIR78k) whereas KNRM and Conv-KNRM have better performances on noisy queries ($+26.56\%$ and $+5.32\%$, respectively on the nDCG@5 compared to wikIR78k). Moreover, with the exception of KNRM, all models designed specifically for \textit{ad-hoc} IR perform better than BM25 on all metrics with statistical significance. However DRMM does not achieve the best performances anymore. This indicates that DRMM is best suited for short and well defined queries but other models with more parameters such as Conv-KNRM and DUET are more robust to noise given enough training data.\\

\section{Conclusions and future work}
In this paper, we propose \textit{WIKIR} a toolkit for building large-scale English information retrieval dataset from \textit{Wikipedia}. \textit{WIKIR} can also be used to train and evaluate deep text matching models. We propose a general framework to construct an IR dataset from any resource that satisfies three topical properties. Additionally, we made available for download wikIR78k and wikIRS78k: two large-scale IR datasets built using \textit{WIKIR}, that are well suited for designing and training deep models for \textit{ad-hoc} IR. All our code is available and our experiments are reproducible.

For future work, we plan to use wikIR78k and wikIRS78k to pre-train deep models for \textit{ad-hoc} IR and fine-tune them on standard IR datasets to see if any gain is obtained compared to weak supervision~\cite{46572}. We will also adapt \textit{WIKIR} to more languages and try our framework to produce IR datasets from other resources such as PubMed Central.\footnote{\url{https://www.ncbi.nlm.nih.gov/pmc/}}

\section{Acknowledgements}
The authors would like to thank Maximin Coavoux,\cref{lig_uga} Emmanuelle Esperança-Rodier,\cref{lig_uga} Lorraine Goeuriot,\footnote{\label{lig_uga}LIG, Université Grenoble-Alpes} William N. Havard,\cref{lig_uga}
Quentin Legros,\footnote{School of Engineering and Physical Sciences, Heriot-Watt University, Edinburgh} Fabien Ringeval,\cref{lig_uga} and Loïc Vial\cref{lig_uga} for their thoughtful comments and efforts towards improving our manuscript.

\section{Bibliographical References}
\label{main:ref}

\bibliographystyle{lrec}
\bibliography{main}


\end{document}